\documentclass{emulateapj}
\usepackage[]{graphicx}

\shorttitle{Molecular Outflows and Abundances in the Local ULIRG Population}
\shortauthors{Chung et al.}
\begin{document}

\title{Evidence for 1000 km\,s$^{-1}$ Molecular Outflows in the Local ULIRG Population}

\author{Aeree Chung\altaffilmark{1,2}, Min S. Yun\altaffilmark{3}, 
        Gopal Naraynan\altaffilmark{3}, Mark Heyer\altaffilmark{3},
        Neal R. Erickson\altaffilmark{3}}

\altaffiltext{1}{Department of Astronomy, Yonsei University, Seoul 120-749, 
                 Republic of Korea; email: achung@yonsei.ac.kr}
\altaffiltext{2}{Harvard-Smithsonian Astrophysical Observatory, 
                 60 Garden Street, Cambridge, MA 02138}
\altaffiltext{3}{Department of Astronomy, University of Massachusetts, 
                 710 North Pleasant Street, Amherst, MA 01003, USA}

\begin{abstract}
The feedback from galactic outflows is thought to play an important role in shaping the gas content, star formation history, and ultimately the stellar mass function of galaxies.   Here we present evidence for massive molecular outflows associated with ultra-luminous infrared galaxies (ULIRGs) in the coadded Redshift Search Receiver $^{12}$CO (1--0) spectrum.  Our stacked spectrum of 27 ULIRGs at $z=0.043-0.11$ ($\nu_{\rm rest}=110-120~$GHz) shows broad wings around the CO line with $\Delta V(FWZI)\approx$ 2000~km~s$^{-1}$.  Its integrated line flux accounts for up to 25$\pm5$\% of the total CO line luminosity.  When interpreted as a massive molecular outflow wind, the associated mechanical energy can be explained by a concentrated starburst with $SFR\ge100~M_\odot~$ yr$^{-1}$, which agrees well with their $SFR$ derived from the $FIR$ luminosity.
Using the high signal-to-noise stacked composite spectrum, we also probe $^{13}$CO and $^{12}$CN emission in the sample and discuss how the chemical abundance of molecular gas may vary depending on the physical conditions of the nuclear region.
\end{abstract}
  
\keywords{galaxies: evolution --- galaxies: interactions --- galaxies: ISM --- galaxies: nuclei  --- galaxies: starburst --- ISM: jets and outflows }

\section{INTRODUCTION}
\label{sec-intro}
Extreme star formation rates of $\gtrsim10^3~M_\odot$~yr$^{-1}$ have been derived for luminous infrared galaxies discovered by deep $IR$ surveys \citep{ds98,younger08,riechers09}. In such radiation-pressure supported galactic disks with the maximum starburst \citep{thompson05}, large scale outflows can be triggered by superwinds from massive young stars and supernova explosions, which may play important roles in galaxy formation and evolution \citet{cooper08}. Galactic outflows can regulate star formation by heating cool gas \citep{tang09}. They can enrich both intergalactic medium and galactic disks \citep{heckman90}. Their feedback can also explain the apparent discrepancy between the theoretical prediction of the dark matter halo mass function and the measured stellar mass function for galaxies in the successful $\Lambda$CDM scenario \citep{sh03}. 

In this Letter, we probe kinematic signatures of molecular outflows in a sample of 27 ultra-luminous infrared galaxies (ULIRGs) recently studied by \citet{chung09}.  A large fraction of ULIRGs are mainly powered by merger-induced starbursts \citep{sanders88} and hence make good targets for investigating associated outflows. Evidence for such an outflow has been reported in a single ULIRG system such as Arp~220 \citep{sakamoto09} and Mrk~231 \citep{feruglio10}. Those outflow signatures are however too faint to be  detected individually in our ULIRG sample. Therefore we employ a stacking analysis to look for faint and broad high velocity line wings in the $^{12}$CO line profile. In order to inspect outflow driving mechanisms, our ULIRG sample is partitioned into two groups based on optical emission line diagnostics: starburst dominated galaxies and galaxies with large AGN contributions. With the reduced noise in the stacked composite spectrum, we also measure the average brightness of other weaker molecular lines such as $^{13}$CO(1--0) and $^{12}$CN(1--0), as an independent measurement of molecular gas properties, that have been detected only in the nearest $IR$ luminous galaxies \citep[e.g.][]{aalto95,aalto02}. 


\section{SAMPLE and STACKING}
\label{sec-sample}

We use the sample and the data from the recent Redshift Search Receiver
(RSR) $^{12}$CO $J=1\rightarrow0$ survey of local ULIRGs by \citet{chung09} .
The observations were carried out with the Five College Radio Astronomy Observatory (FCRAO) 14m Telescope in 2007 and 2008, targeting 29 ULIRGs at $z=0.043-0.11$. As discussed in detail by \citet{chung09}, this is a representative subset of ULIRGs as the primary selection criteria were those related to observational scheduling and the redshift range that brings the $^{12}$CO  $J=1\rightarrow 0$ line within the bandpass of the RSR system.
In our stacking analysis, we include only the 27 CO detected objects.
The CO line luminosity $L_{\rm co}^\prime$ of the sample ranges from 1.2 to 
15.3$\times10^9~$K~km~s$^{-1}$~pc$^{2}$ with a median value of 6.7$\times10^9~$K~km~s$^{-1}$~pc$^{2}$.

The stacking of the RSR spectra is performed using the following procedure. First, each coadded spectrum is shifted to the rest frequency by multiplying the observed frequency by (1+$z_{\rm co}$), where $z_{\rm co}$ is the CO redshift of each ULIRG derived from the line fitting \citep{chung09}.  Each CO spectrum is normalized by the best fit Gaussian peak, and then all spectra are {\it ``aligned''} at the frequency centroid.  A linear interpolation is used in the alignment process.   The normalized spectra are averaged, weighted by the $rms$ ``noise'' measured in the normalized spectra, excluding the $\pm0.5~$GHz regions around the three transitions of our interest, $^{13}$CO, $^{12}$CN, and $^{12}$CO as well as the noisy end channels. Finally, the averaged spectrum is Hanning smoothed to produce the final spectral resolution of 61~MHz (158~km~s$^{-1}$ at 115.27~GHz). 

A ``non-ULIRG'' comparison spectrum was derived by stacking the RSR spectra of 19 $z=0.037-0.066$ galaxies selected for their high HI mass ($M_{\rm HI}\gtrsim2\times10^{10}~M_\odot$; Haynes et al. in prep., O'Neil et al., in prep).  These 19 galaxies were selected from another RSR commissioning programs - the CO survey of 29 HI rich galaxies at similar redshifts.  Nineteen galaxies were detected in CO with comparable $S/N$ and sensitivity as our ULIRG sample, but they are otherwise normal in their star formation and nuclear activities. These HI rich galaxies mostly look like normal spirals in the optical and their mean $FIR$ luminosity is $2.8\pm1.4\times10^{10}~L_\odot$, 30 times lower than that of our ULIRG sample.
Their $L_{\rm co}^{\prime}$ of 0.4--3.2$\times10^9$~K~km~s$^{-1}$~pc$^{2}$ is only slightly smaller than that of our ULIRG sample. 
Further details of the RSR CO observations of these HI-rich spirals will be presented elsewhere (Chung et al. in prep).

\begin{figure}
\plotone{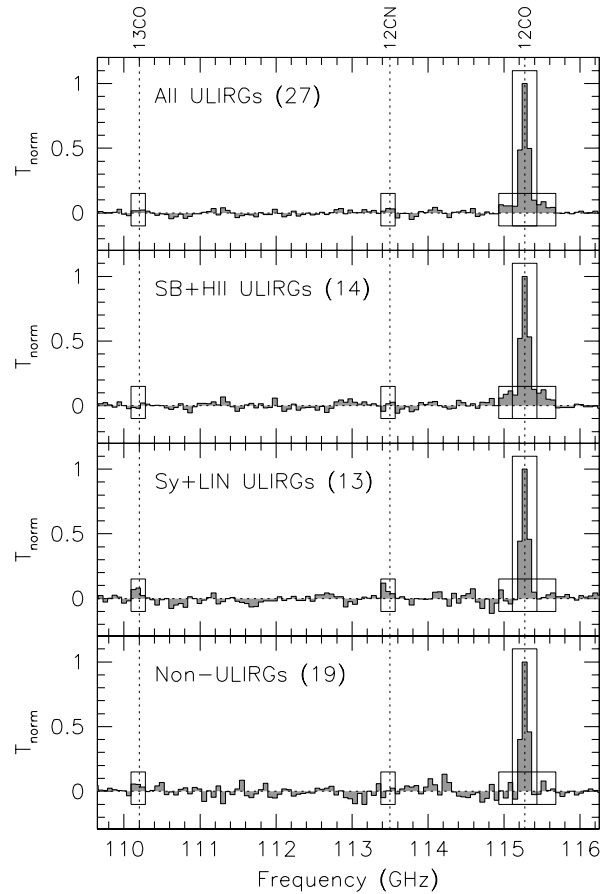}
\caption{Full Composite Redshift Search Receiver (RSR) spectra of the ULIRG and comparison sample.
The rest frequencies of $^{13}$CO (110.20~GHz), $^{12}$CN (113.50~GHz), and $^{12}$CO (115.27~GHz) are indicated with dotted vertical lines. 
The stacked spectrum of 27 $^{12}$CO detected ULIRGs is shown on the top, the stacked spectrum of a subsample of 14 ``Sbrst'' or ``H{\sc ii}'' (SB group, $L_{IR}^{\rm AGN}\lesssim0.12L_{IR}$) and 13 seyfert or LINER type ULIRGs (AGN group, $L_{IR}^{\rm AGN}\approx0.32L_{IR}$) are shown in the middle, followed by the stacked spectrum of 19 non-ULIRG sample at the bottom. The FWZI spectral regions used to measure the line and wing flux density are shown as boxes.
\label{fig-chemi}}
\end{figure}

The rms noise in the final spectra for the ULIRG sample and the control sample, normalized by the $^{12}$CO peak flux, are 0.014 and 0.027, respectively, yielding the S/N that is better than those of individual spectra by a factor of 5 to 47 (see Table~\ref{tbl-stack}).  Figure~\ref{fig-chemi} shows the stacked, normalized spectrum of a broad frequency range (109.65--116.25~GHz) which includes all three transitions, $^{13}$CO (1--0), $^{12}$CN (1--0), and $^{12}$CO (1--0).  In Figure~\ref{fig-stack}, we zoom in the 4500~km~s$^{-1}$ range around the $^{12}$CO line to show the characteristics of the profile more in detail.

\section{STARBURST POWERED OUTFLOWS}
\label{sec-outflow}
\begin{figure}
\plotone{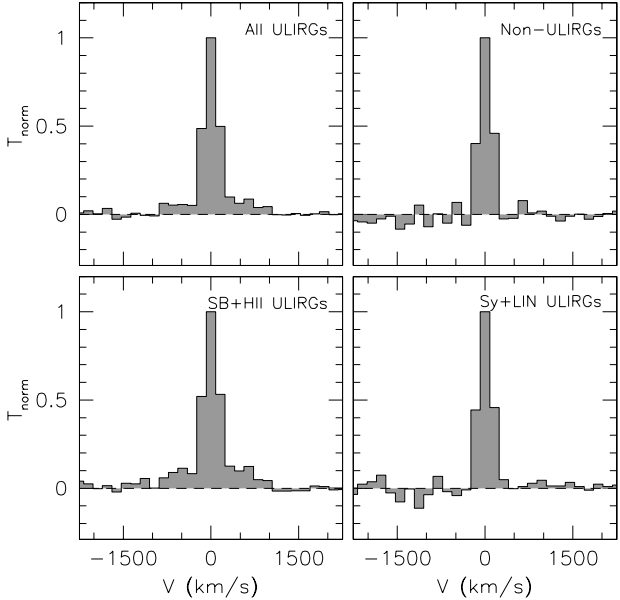}
\caption{A zoom-in view of the composite $^{12}$CO $J=1\rightarrow0$ Redshift Search Receiver (RSR) spectra. On the top, the ULIRG sample of 27 CO detected galaxies from \citet[][]{chung09} is compared to the non ULIRG sample of 19 HI-rich galaxies. On the bottom, the CO spectra of the SB group and the AGN group are compared.\label{fig-stack}}
\end{figure}

The stacked spectrum of the ULIRG group has revealed broad wings around the CO line as seen on the top of Figure~\ref{fig-chemi}.. 
The wings are blue- and redshifted by $\approx1000~$km~s$^{-1}$ from the main CO line peak (FWZI of 2000~km~s$^{-1}$) with the total line integral of 19$\pm5$\% of the total (see Fig.~\ref{fig-stack}).  The line wings of the ULIRG stacked spectrum are detected with S/N$\sim3$ in each channel, which would be difficult to be detected in individual spectra.
In fact, the effective integration time of the stacked ULIRG spectrum is 115.8~hrs, 10--60 times more than the integration on individual ULIRGs. 
Note that such wings are not present in the control sample of HI rich galaxies with the effective integration 127.7~hrs as shown on the bottom of
Figure~\ref{fig-chemi}. The comparison of the ULIRGs with the non-ULIRG population is better shown in the upper two panels of Figure~\ref{fig-stack}.

\begin{table*}
\centering
\scriptsize
\caption{Outflow Properties and Molecular Abundance\label{tbl-stack}}
\begin{tabular}{rccccccc}
\hline\hline
\multicolumn{1}{l}{Group}  
&\multicolumn{4}{c}{----------------------------- Mean -----------------------------}
&\multicolumn{3}{c}{------------ Stacked ------------}\\
&$z_{\rm co}$ 
&rms (mK/$T_A^*$)
&$T_{\rm peak}$/rms
&$W_{\rm co}$ (km/s)
&$\frac{{\rm wing}}{{\rm 12CO}}$ 
&$\frac{{\rm 12CO}}{{\rm 13CO}}$ 
&$\frac{{\rm 12CO}}{{\rm 12CN}}$\\
\hline
\multicolumn{1}{l}{ULIRGs................}&&&&&&&\\
 All (27)              &0.072$\pm$0.023& 0.46$\pm$0.15&6.4$\pm$4.1& 263$\pm$~59&0.19$\pm$0.05 &  $\geq$~16.6 & $\geq$~16.6\\
 Sbrst+H{\sc ii} (14)  &0.071$\pm$0.027& 0.45$\pm$0.15&7.3$\pm$4.2& 266$\pm$~67&0.25$\pm$0.06 &  $\geq$~13.3 & $\geq$~13.3\\
 Sy+LIN (13)           &0.073$\pm$0.026& 0.47$\pm$0.16&5.4$\pm$4.0& 262$\pm$~54&$\leq$~0.12   & 11.1$\pm$6.2 & $9.3\pm4.3$\\
\multicolumn{1}{l}{Non-ULIRGs.........~~}&&&&&&&\\	              	  
 HI-rich spirals (19)  &0.050$\pm$0.007& 0.39$\pm$0.09&5.0$\pm$2.7& 266$\pm$103&$\leq$~0.11   &  $\geq$~~7.3 & $\geq$~7.3\\
\hline
\end{tabular}
\end{table*}

Such broad wings can form when entrained cool gas gets ejected along with hot ionized outflowing gas \citep{curran99,narayanan06}.
In order to examine whether a starburst or an AGN is powering the outflow, we have divided the ULIRG sample into two groups: (1) 14 ULIRGs which are classified as ``Sbrst'' or ``H{\sc ii}'' (SB group) with no obvious sign of AGNs; and (2) 13 ULIRGs with ``seyfert'' spectra (AGN group). This grouping is done based on the classification from the NASA/IPAC Extragalactic Database (NED)\footnote{This research has made use of the NASA/IPAC Extragalactic Database (NED) which is operated by the Jet Propulsion Laboratory, California Institute of Technology, under contract with the National Aeronautics and Space Administration.}.  
Objects with a ``LINER'' classification are included in the AGN group unless it also has a ``Sbrst'' or ``H{\sc ii}'' designation. Galaxies with a hybrid (``Sy+SB/H{\sc ii}'') classification are included in the AGN group. Among our sample, eight ULIRGs from each group have been modeled in their spectral energy distribution by \citet{farrah03} who found an AGN contribution to the $IR$ luminosity of $\ge$27\% for the eight in the AGN group, which is a factor of two higher than that of the eight in the SB group ($\sim12$\%). The mean S/N measured by the ratio of the CO line peak and the rms is highest for the SB group due to a few objects with the strongest CO emission among the sample, but this ratio does not vary significantly from group to group. In fact, the mean rms of each group and the CO linewidths of different groups are very similar as summarized in Table~\ref{tbl-stack}, making our results robust.

We show the comparison of the two groups on the bottom of Figure~\ref{fig-stack}. The rms noise in the normalized stacked spectra is 0.0189 and 0.032 for the SB group and the AGN group, respectively. The wings around $^{12}$CO line appears to be even stronger in the SB group compared to the wings seen in the entire ULIRG sample, with more flux in the wings which is about 25\% of the total CO flux. These broad features however disappear when only the ULIRGs with Seyfert spectra are combined. The total line flux was measured by integrating the line flux density within the FWZI regions (thin solid lines in Fig. 1) for both groups with and without wings.  The wing-only flux was measured within the same FWZI as the AGN group and the control sample, and the difference between the total and the line wing flux has been adopted as the CO flux for these groups.  The fractional flux in the wings and the upper limits of different subgroups are summarized in Table~\ref{tbl-stack}.
\citet{rvs05} also found a lower frequency of neutral wind among Seyfert 2 ULIRGs in their study of Na~I~D absorption line in 26 AGN/starburst-composite ULIRGs at $0.03<z<0.44$, further supporting the starburst origin for this massive neutral wind.

The energetics of the neutral wind traced in broad CO wings is also consistent with being powered by the ongoing starburst traced in the far-infrared. The energy injection rate ($dE/dt$) in a wind-blown bubble of a radius $R$ expanding at velocity $v$ into an infinite homogeneous medium with density $n_0$ can be expressed as \citep{weaver77},
\begin{equation}
\frac{dE}{dt}\sim3.3\times10^{35}R_{kpc}^2v_{km/s}^3n_{0,cm^{-3}}~erg~s^{-1}.
\end{equation}
Adopting a bubble size of $0.2~$kpc \citet{sakamoto09} found for the high velocity molecular wind in Arp~220, an outflow velocity of $1000$~km~s$^{-1}$ from the line wing velocity in the stacked spectrum, and an ambient density of 10~cm$^{-3}$ \citep[e.g.][]{veilleux95}, we drive an energy injection rate of $\sim1.3\times10^{44}$~ergs~s$^{-1}$. Assuming an energy output per supernova of $\sim10^{51}$~ergs \citep{veilleux95}, our estimated $dE/dt$ yields 
a supernova rate, $\nu_{\rm SN,yr^{-1}}$ of $4$~yr$^{-1}$. Using a Scalo initial mass function (IMF) with a mass range of $5-100~M_\odot$, the star formation rate inferred from this supernova rate \citep[$SFR_{M>5M_\odot}=24.4~\nu_{\rm SN,yr^{-1}}~M_\odot$~yr$^{-1}$,][]{condon92,rosa05} is $SFR\approx100~M_\odot~$yr$^{-1}$.  This agrees well with the $SFR$ derived from the $FIR$ luminosity for these ULIRGs, 134-352~$M_\odot~$yr$^{-1}$ \citep[see Eq. 8 of][]{hopkins03}.

The outflow speed implied by the CO line wings, $1000$~km~s$^{-1}$, is comparable to the wind velocity measured by other phases of the superwind such as H$\alpha$ and Na~D \citep[400-800~km~s$^{-1}$,][]{martin99} and OH
\citep[1400~km~s$^{-1}$,][]{fischer10}. If they all trace the same outflow, then Eq. (1) suggests that the spatial extent of Na D and H$\alpha$ should be larger ($\sim0.4~$kpc) while the spatial scale for the OH winds measured by Herschel would be more compact ($\lesssim0.1$~kpc) than the CO wind. The fact that wind velocities measured in molecular outflows are larger than in the optical may indicate that supernova driven wind embedded in molecular gas slows down by the time it breaks out of the starburst region. 

The inferred outflow speed exceeds the escape velocity and is high enough to blow away the molecular gas that hosts the star formation activity and to pollute the surrounding IGM significantly.  Depending on whether the CO line is optically thin or thick, the outflowing molecular gas mass ranges between 1-6$\times10^9~M_\odot$.
The line wings are symmetric in intensity and shape on both sides of the line,
and this suggests that the wings are bipolar in geometry and likely optically thin. This is not contradictory to the highly asymmetric CO $3\rightarrow2$ line wings in Arp~220 (FWZI = 1000~km~s$^{-1}$) found by \citet{sakamoto09} since the $J=3\rightarrow2$ line has a higher optical depth. These observations suggest that more than $10^9~M_\odot$ of molecular gas can be removed from the central starburst region through such a wind, rapidly depleting the gas supply for the starburst.  
Some of this gas may eventually rain back onto the galaxy, enriching the galactic disk \citep{heckman03}.

Our conclusion that the central starburst can power the observed massive outflow contradicts the conclusion by \citet{feruglio10} that the 750~km~s$^{-1}$ wind in CO $1\rightarrow0$ found in Mrk~231 is powered by the AGN activity and thus is an example of a ``quasar feedback'' at work.  Although \citet{feruglio10} adopted a smaller than Galactic CO-to-H$_2$ conversion factor, they may still have over-estimated gas mass and mass outflow rate as an optically thin estimate leads to a $\ge2$ times smaller gas mass, lowering the mass outflow rate much closer to the current $SFR$.  A stronger case for an AGN-driven molecular outflow is found in NGC~1266 where the optically thin CO mass outflow rate clearly exceeds the observed current $SFR$ (Alatalo et al., in prep).  The CO outflow velocity is much lower ($\sim400$~km~s$^{-1}$), however, and this phenomenon may not be very common, at least at low-$z$, since this is the only object with an AGN-driven molecular outflow found in their survey of a large number of early-type galaxies.  \citet{narayanan08} have shown that an AGN-driven molecular outflow may persist longer than a SB-driven outflow using a numerical simulation, but the accuracy of such model predictions and the sub-grid physics included needs to be tested further using a large sample of AGN+SB systems.

\section{13CO AND 12CN ABUNDANCES}
\label{sec-abundance}
Two important molecular transitions also appear in our stacked composite spectra, and we examine their line strengths to gain further insights into the molecular ISM in these ULIRGs. 
Those are the lowest transitions of $^{13}$CO and $^{12}$CN (CN hereafter) at 110.20 and 113.50~GHz, which have been detected in local starburst and 
Seyfert galaxies \citep{casoli92,aalto95,aalto02,perez07}.

In the RSR composite spectra, we find $\gtrsim3~\sigma$ bumps at both $^{13}$CO and CN frequencies only in the AGN ULIRG group with the flux ratios, 
$11.1\pm6.2$ and  $9.3\pm4.3$ for $^{12}$CO/$^{13}$CO and 
$^{12}$CO/CN, respectively (Fig.~\ref{fig-chemi}). The other groups do not show such features, and the lower limits in $^{12}$CO/$^{13}$CO and 
$^{12}$CO/CN are summarized in Table~\ref{tbl-stack}. The same linewidths as the AGN group have been adopted to calculate the upper limits of $^{13}$CO (660~km~s$^{-1}$) and CN (600~km~s$^{-1}$) for the other groups.

The ratio $^{12}$CO/$^{13}$CO ($R_{10}$) has been reported to be generally larger in starburst galaxies \citep[$R_{10}\geq20$,][]{gh01} than in optically thick normal spirals \citep[$10<R_{10}<20$,][]{casoli92} or Seyfert galaxies \citep[$R_{10}\approx12$,][]{ps98}. It has been suggested that the overproduction of $^{12}$C, a primary product of nucleosynthesis \citep{bb96} in actively star forming galaxies, is responsible for this trend \citep{casoli92}. Alternatively,
\citet[][also 1995]{aalto91} have suggested that $R_{10}$, which gauges the optical depth of $^{13}$CO gas in LTE \citep[$I_{\rm^{12}CO}/I_{\rm^{13}CO}\approx1/\tau_{\rm^{13}CO}$;][]{paglione01}, can increase when 
molecular clouds are disturbed by powerful tidal force in merger driven starburst galaxies.  Increased velocity dispersion within GMCs and a broader cloud-to-cloud velocity distribution can reduce the $^{13}$CO opacity within these starburst nuclei. 
Meanwhile, CN is known to be a tracer of dense gas with lower critical density than HCN \citep[by a factor of 5;][]{perez07,baan08}.  CN molecule is a photo- or X-ray dissociation product of HCN and HNC \citep{baan08}, and is predicted to be abundant in both PDRs and XDRs \citep{kohno08}.  \citet{meijerink07} however found the CN/HCN ratio to be enhanced in XDR than in PDR, and toward the DR edges where the gas is highly ionized as in XDR. Our finding of the lowest CO/CN ratio in the AGN ULIRG group may imply
a higher ionization rate of dense molecular gas as predicted by the Meijerink et al. models. 

We have only 8 (four SB and four AGN) and 14 objects (nine SB and five AGN) whose $^{13}$CO and CN lines fall within the RSR frequency band, and the significance of these results will have to be confirmed with a larger sample.

\section{FUTURE PROSPECTS}
\label{sec-future}
There are ongoing theoretical efforts to model galactic outflows to understand their detailed properties such as their frequency and energetics \citep[e.g.][]{cn10} and the feedback on scaling relations of galaxies 
\citep[e.g.][]{sales10}.  Even for objects at cosmological distances where more direct morphological clues such as superbubbles, filaments, and chimneys are not visible, these outflow models can be tested by examining their spectroscopic signatures.  Presently there is no consensus to on the driving mechanism for the observed outflows: \citet[i.e. starburst - ][vs. AGN - Ferglio et al. 2010; Fischer et al. 2010]{sakamoto09,riechers09}.  Obtaining a better understanding is a pre-requisite in evaluating the importance of outflow feedback plays in galaxy evolution.  The Redshift Search Receiver on the Large Millimeter Telescope (LMT) - a 50m single-dish facility being built at Volc$\acute{\rm a}$n Sierra Negra, near Puebla, Mexico will extend our capability to study galaxy outflows and winds at higher redshifts with its vastly improved sensitivity.
Spatially resolved morphological and kinematical details obtainable using the ALMA will offer us the most stringent observational test for the origin of these massive molecular outflows.

We are grateful to Mike Brewer, Don Lydon, Kamal Souccar, Gary Wallace, Ron Grosslein, John Wielgus, Vern Fath, and Ronna Erickson for their technical support of the Redshift Search Receiver commissioning.  This work was supported by NSF grants AST 0096854, AST 0540852, and AST 0704966.  We also thank K. Alatalo, D. Sanders, and N. Scoville for their helpful discussions. Support for this work was (also) provided by the National Research Foundation of Korea to the Center for Galaxy Evolution Research.

\end{document}